\documentclass[10pt,conference,letterpaper]{IEEEtran}
\usepackage{times,amsmath,epsfig}

\usepackage{cite} 
\usepackage{graphicx}
\usepackage{xcolor}
\usepackage{xspace} 

\usepackage{caption}
 
\usepackage{multirow}

\usepackage{pifont}

\usepackage{arydshln}

\usepackage{paralist}

\usepackage[T1]{fontenc}

\usepackage{wrapfig}

\newcommand{\project}{\textit{InfiniViz}\xspace}

\renewcommand{\paragraph}[1]{\noindent\textbf{#1:}}

\title{InfiniViz: Interactive Visual Exploration using Progressive Bin Refinement}
\author{%
{Niranjan Kamat{\small $~^{1}$}, Arnab Nandi{\small $~^{2}$} }%
\vspace{1.6mm}\\
\fontsize{10}{10}\selectfont\itshape
Computer Science and Engineering Department, The Ohio State University\\
2015 Neil Avenue, Columbus, OH 43210, USA\\
\fontsize{9}{9}\selectfont\ttfamily\upshape
\{kamat.14$^{1}$, nandi.9$^{2}$\}@osu.edu

}

\begin{document}
%

\newcommand{\cblue}{\textcolor{blue}}

\newcommand{\cmark}{\ding{51}}%
\newcommand{\xmark}{\ding{55}}%
                     
\maketitle

\begin{abstract}

Interactive visualizations can accelerate the data analysis loop through near-instantaneous feedback.
To achieve interactivity, techniques such as data cubes and sampling are typically employed.
While data cubes can speedup querying for moderate-sized datasets, they are ineffective at doing so at a larger scales due to the size of the materialized data cubes.
On the other hand, while sampling can help scale to large datasets, it adds sampling error and the associated issues into the process.

While increasing accuracy by looking at more data may sometimes be valuable, providing result minutiae might not be necessary if they do not impart additional significant information.
Indeed, such details not only incur a higher \emph{computational} cost, but also tax the \emph{cognitive} load of the analyst with worthless trivia.
To reduce both the computational and cognitive expenses, we introduce \project{}. Through a novel result refinement-based querying paradigm, \project{}
provides error-free results for large datasets by increasing bin resolutions progressively over time.
Through real and simulated workloads over real and benchmark datasets, we evaluate and demonstrate \project{}'s utility at reducing both cognitive and computational costs, while minimizing information loss.

\end{abstract}

\section{Introduction}




Visualizations are widely used in data analysis.
In this era of \emph{Big Data}, querying large datasets has become a necessity. 
While analyzing large datasets helps discover insights that are otherwise unattainable~\cite{manyika2011big, provost2013data, chen2012business}, querying such large datasets is computationally expensive and inconducive to interactivity. 
Providing results within interactive latencies~($<500$ms) has been shown to greatly benefit  analysis, with failing to do so having significant adverse consequences on the analysis outcomes~\cite{shneiderman1984response, liu2014effects}.
Marrying these twin concerns of interactivity and the need to query large datasets presents us with two compelling, contradictory forces.

A popular technique to reconcile them is to only process a sample of the data.
However, this necessitates providing not only the query result but also its error~\cite{ olken1993random, chaudhuri1999random, agarwal2013blinkdb, kandula2016quickr}. 
Interpretation of sampling error, even by experts, has been known to be error-prone~\cite{belia2005researchers, cumming2013understanding}. 
Annotation of visual results with errors can further introduce clutter~\cite{ferreira2014sample, olston2002visualizing, moritz2017trust}.

\begin{figure}[t]
   \centering
   \captionsetup{justification=centering}
   \includegraphics[width=\columnwidth]{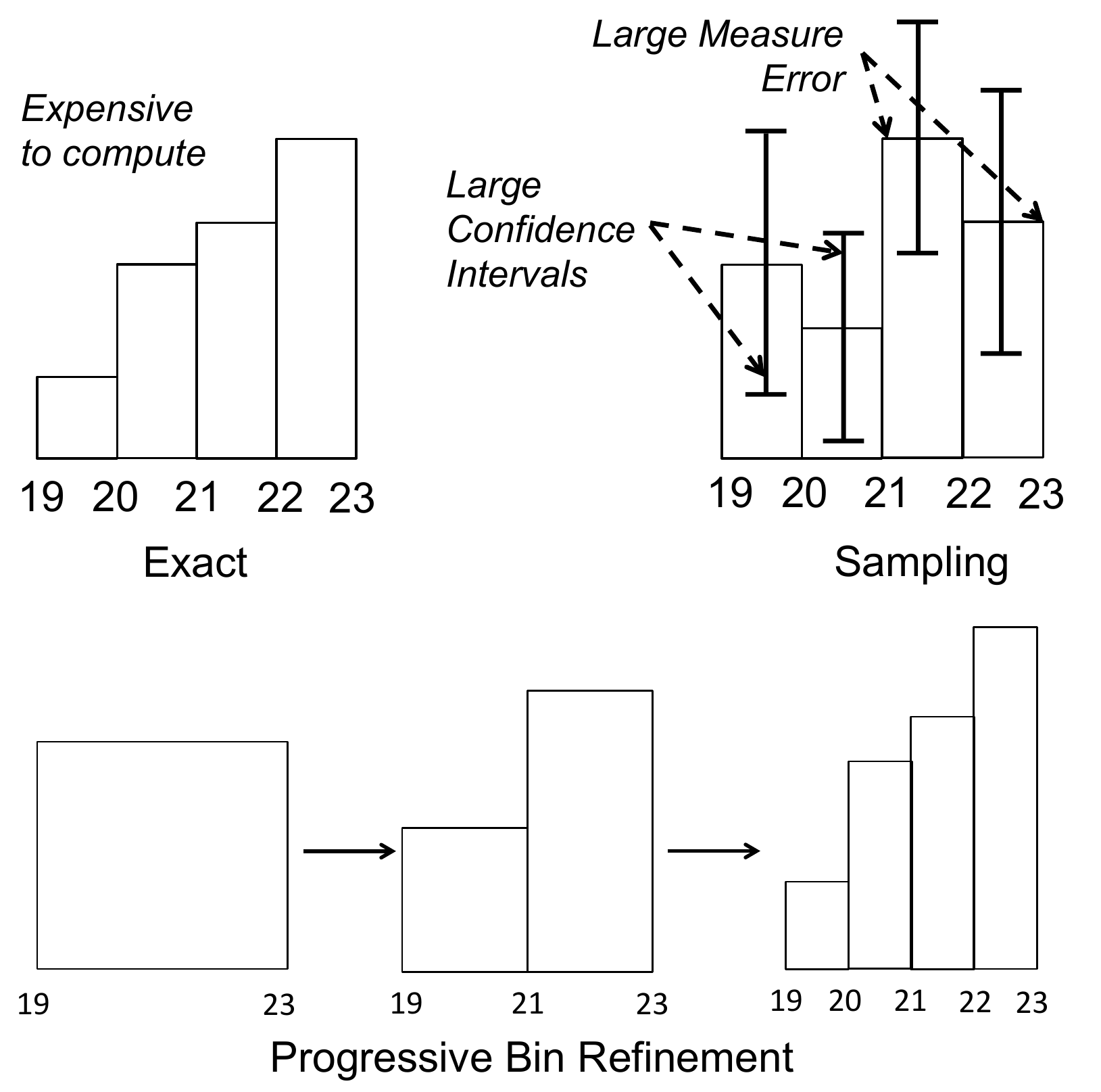}
   \vspace{-10pt}
   \caption{Approximate Querying Modes: While sampling and online aggregation can result in large sampling error, especially during interactive response times, \emph{progressive refinement} can deliver results without measure error from the get-go. The result resolution can be increased over time.}
   \label{fig:approximate_querying_options}
   \vspace{-20pt}
\end{figure}

\nocite{haas1996selectivity, haas1996hoeffding}
Online aggregation builds on sampling by accessing more data over time, thereby reducing the error~\cite{hellerstein1997online, haas1999ripple,  jermaine2005disk, nirkhiwale2013sampling, li2016wander}.
However, error-free results are unavailable till the entire dataset is processed.
Further, as sampling error depends on the quality and size of the sample, it is common for sampling error to be large, especially at lower sampling rates, which can be expected during interactive response times~(Figure~\ref{fig:approximate_querying_options}). 
Additionally, highly selective queries reduce the number of tuples passing through, thereby lowering the effective sampling rate and increasing the error.
Data skew further worsens these issues.


Another common technique used to achieve interactivity is data cubes~\cite{gray1997data} -- a  cube contains pre-computed aggregates for user-specified measures for all possible column combinations. 
Consequently, user queries can be run over the pre-computed result sets, which are usually  smaller by multiple orders of magnitude.
These result sets can be indexed, compounding the query speedups.
However, cube size increases exponentially with the number of columns and their cardinalities, increasing the cube materialization cost, but 
more importantly from an interactive querying perspective, 
the time needed to query it -- 
a cube constructed over a dataset with $m$ dimensions, with the $i^{th}$ dimension having cardinality $d_i$, can consist of up to $\prod_{i = 1} ^ {m} d_i$ rows for each measure. 
Thus, constructing cubes over large datasets is inconducive to the pursuit of interactivity. 

\begin{figure*}[t]
	\centering
   \captionsetup{justification=centering}
   \includegraphics[width=0.8\textwidth]{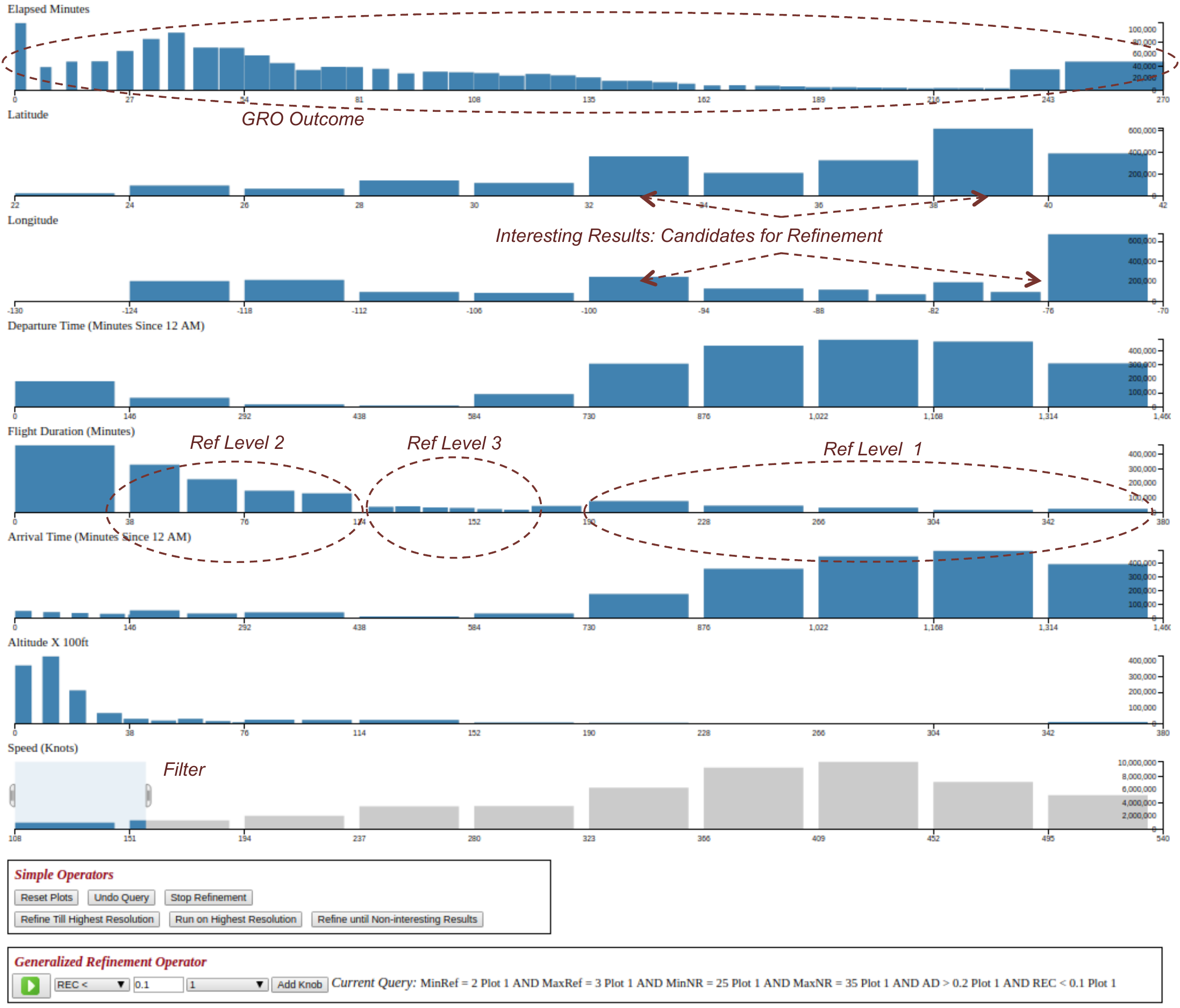}
      \caption{\project{} during \emph{Progressive Refinement}: We see a user's view in her interactive exploration of a dataset having 50M rows and 8 columns, \emph{and importantly 8 linked views}, through filter and refine queries~(Section~\ref{cog_com:actions_behavior}).}
   \label{fig:screenshot}
	 \vspace{-15pt}
\end{figure*}



In the context of visualizations, however, constructing a cube over the entire dataset might not be necessary.
As screen resolution limits the information that can be presented to the user, data binning is a natural consequence, and has been looked at previously by multiple systems including Profiler~\cite{kandel2012profiler}, imMens~\cite{liu2013immens}, and Nanocubes~\cite{lins2013nanocubes}, which construct cubes over the smaller binned datasets. 
This reduces the size of the data cube that needs to be computed, thus enabling interactive query execution.
These systems also allow a user to increase the refinement~(zoom) level of a result, thereby providing finer-grained results due to the smaller bin sizes. 

In this paper, we run with this concept of result refinement, delving into its multiple benefits in delivering approximate visualizations whose resolutions increase over time.
This leads us to propose a novel querying paradigm  
-- \emph{progressive refinement}~(Section~\ref{prog_refine}).
We treat result refinement as one of the primary query operators, alongside filtering~(Section~\ref{cog_com:actions_behavior}).
While being mindful of the required interactive latency, interesting results are refined over time to increase the resolution of the results.

Previous binning-based systems usually allow a user to refine the results by letting her specify the refinement level.
We extend this approach by introducing a generalized, richer refinement operator, that allows for specification of multiple refinement related criteria such as the number of results, average deviance, relative entropy change, in addition to the refinement level~(Section~\ref{cog_com:knobs}).
As these criteria might be contradictory with each other (consider maximum number of results vs minimum refinement level), we use the result content to trade them off through our novel information theory-based metrics~(Section~\ref{doa:result-ranking}), which results in our \emph{Generalized Refinement Operator~(GRO)}~(Section~\ref{doa:auto-zooming}).
While other systems have used binning as a means to achieve data reduction, we look at binning through the lens of approximate querying, including the \emph{notion of error}~(Section~\ref{doa:AD}) in such systems.

Our experiments demonstrate that not only is the \project{} response time low~($< 100$ ms), but the overall computational cost is also a couple of magnitudes lower than the cost of querying the underlying non-binned dataset~(Section~\ref{expt}).
Our detailed user studies demonstrate, through statistically significant results, \project's ability to accelerate not only the individual queries, but more importantly, the overall data analysis loop as well.

Further, not only is the computational cost reduced, but the cognitive load over the user in understanding the results reduces as well.
In analyzing individual interesting results, e.g. comparing multiple bins or figuring out relationship between different distributions, it is known that having numerous uninteresting results can hurt the analysis~\cite{ariely2001seeing, alvarez2009spatial}.
Indeed, our user studies also demonstrate that deluging the user with insignificant results hinders her analysis.
As \project{} does not inundate the user with multiple insignificant results, she can focus on the interesting results.
Further, as people excel at summarizing and generating patterns from visual data, having fewer results does not hurt in this endeavor~\cite{keim2008visual, simakov2008summarizing}.


Our progressive refinement approach can be summarized by the following SQL query. Note that while we consider the use-case of histograms, our approach can be extended in a straightforward fashion to heatmaps as well.

\vspace{2pt}
\noindent \texttt{SELECT agg\_func(agg\_col) AS y\\
FROM UNION(set\_of\_binned\_tables)\\
WHERE filters\\
GROUP BY grouping\_col AS x\\
HAVING resolution(x)}

\subsection{Motivating Example}
Let us join Sia, a data analyst, in her exploratory analysis of flight patterns. 
The dataset~(\emph{PAD}, Section~\ref{expt:datasets}) consists of details of individual flights such as their duration, delays in departure and arrival times, distance covered, distance between arrival and destination, etc. 
Our user studies were conducted this dataset, and the user behavior described here resembles that of our users.

Sia wishes to familiarize herself with the data, and extract interesting bits using the standard operators of filter, drill-down, and roll-up.
She wants to explore flight patterns during takeoff and landing. She does so by setting a filter to the \emph{Speed} dimension to only consider flights having low speeds~(Figure~\ref{fig:screenshot}).
She notices that setting the filter changes the \emph{Altitude} plot -- only low altitude results are returned. 
She examines this plot in more detail by clicking on its interesting bars.
The \emph{Elapsed Minutes} dimension is also correlated with the \emph{Speed} dimension, which she refines in an automated fashion through our \emph{GRO} operator, as she wishes to refine the entire \emph{Elapsed Minutes} plot.
Upon observing interesting results in other plots~(\emph{Latitude} and \emph{Longitude}), she might proceed to refine them as well.
She might repeat this process with more \emph{filter} queries followed by \emph{refines}.

\subsection{Contributions}

Thus, we help Sia by making the following contributions:

1. We introduce the concept of \emph{Progressive Refinement} in the context of visualization-based analysis, to provide the user with \emph{error-free} results within interactive response times over large datasets. 
While previous works allow a user to set the refinement level, we treat refinement as a \emph{first-class citizen}, and look at interactive approximate querying through its lens. 
In this endeavor, we provide a novel refinement operator~(\emph{GRO}) that enables multiple resolution-related criteria to be used.

2. We introduce a novel monotonic information-theoretic metric that guides our result refinement approach.

3. Our extensive experiments, using real and simulated workloads over real-world and benchmark datasets, demonstrate not only \project's efficacy at accelerating the data analysis loop, but also validates our proposed refinement-based interactive querying workflow~(Table~\ref{table:misc-userstudy}).

Section~\ref{system} looks at the various concepts that underpin \project{}. We then present its system architecture in Section~\ref{sys_arch}. Section~\ref{expt} empirically validates our approaches. We then look at the related work in Section~\ref{related}, and finally conclude with our parting thoughts in Section~\ref{conclusion}.

\section{InfiniViz System}
\label{system}
We now look at the various concepts underpinning \project. In particular, we further elucidate our \emph{progressive refinement} concept, including its benefits and pitfalls, and look at its parallels with online aggregation. We also describe the possible user actions and the various refinement operators.

\subsection{Progressive Refinement}
\label{prog_refine}
\project{} progressively improves the \emph{result resolution} over time by \emph{refining} the result.
We now formalize these concepts of result resolution and result refinement.

\subsubsection{Result Resolution}
The concept of result resolution is fairly straight-forward -- small bin sizes result in a higher resolution, while larger bin sizes provide lower resolution.

\subsubsection{Result Refinement}
After providing initial results within interactive response times over lower resolution data, the results having higher information loss are refined over time, thereby increasing the result resolution~(Figure~\ref{fig:approximate_querying_options}).

\subsection{User Actions \& Behavior}
\label{cog_com:actions_behavior}
We now look at the various direct and indirect manipulation actions available for interacting with \project{}.
We enable two primary functionalities -- \emph{filter} and \emph{refine}.
A user can specify filters on any visualization~(direct manipulation). This results in a \texttt{WHERE} predicate over the corresponding dimension.
She can refine a specific bin's resolution by clicking it~(direct manipulation).
Other direct manipulation actions include the \emph{Reset Plots} button~(removes all filters), \emph{Stop Refinement} button~(stops any further refinement), and buttons for handy, simple refinement actions~(Section~\ref{handy-operators}).
The generalized refinement operator~(Section~\ref{doa:auto-zooming}) can be used by setting thresholds for the different refinement grammar~(Section~\ref{cog_com:knobs}) knobs~(indirect manipulation).
This operator can be used at the level of all visualizations,
or a single visualization.


\subsection{Refinement Grammar}
\label{cog_com:knobs}
While current binning-based systems
allow a user to set the refinement level, we enrich this traditional approach by providing multiple finely tunable knobs.  
These knobs, which together constitute our refinement grammar, allow a user to indirectly determine the result resolution.

\vspace{2pt}
\subsubsection{Refinement Level}
In a traditional fashion, a user can set the minimum refinement level~(\emph{MinRef}). In addition, she can also set the maximum refinement level~(\emph{MaxRef}).
Their default values are set to 0 and the resolution of the non-binned dataset, respectively.

\vspace{2pt}
\subsubsection{Number of Results}
In addition, a user can specify the minimum~(\emph{MinNR}) or the maximum~(\emph{MaxNR}) number of results to be displayed. 
Their default values are set to 0 and $\infty$, respectively.

\vspace{2pt}
\subsubsection{Average Deviance~(AD)}
\label{doa:AD}
One of the governing principles that any progressive refinement-based system should follow is to prefer refined results that show a marked difference from their expected value -- a sub-bin for a bin $i$ can be considered to impart more information if it deviates significantly from the uniform distribution $Uniform(\frac{y(i)}{|sub-bins(i)|})$,
where $y(i)$ represents the value of the $i^{th}$ bin, and \emph{sub-bins(i)} represents the set of sub-bins of a bin $i$.
This motivates our \emph{AD} metric, which determines how well a bin summarizes it's sub-bins, and can be given for a bin $i$ by 
$\frac{1}{|sub-bins(i)|} \times \sum_{sb \in sub-bins(i)} abs\left( \frac{E(sb) - y(sb)}{E(sb)}\right)$, where $E(sb)$ represents the expected value of a sub-bin $sb$ given its parent bin value.
We summarize \emph{AD} for a plot by the average \emph{AD} of its bins.


We illustrate our metrics using the following running example.
Consider a plot having 4 bins with the y-values $10$, $20$, $30$, and $40$, respectively.
Suppose the individual bins are split into sub-bins having y-values \{$4$, $6$\}, \{$10$, $10$\}, \{$10$, $20$\}, and \{$15$, $25$\}, respectively.
Then, \emph{AD} for the first bin can be given by $\frac{1}{2} \times \left( abs\left( \frac{5 - 4}{5} \right) + abs\left( \frac{5 - 6}{5} \right) \right) = 0.2$. Similarly, \emph{AD} for the other bins will respectively be $0$, $0.33$, and $0.25$. 
\emph{AD} for the plot will be $\frac{0.2 + 0 + 0.33 + 0.25}{4} = 0.19$.

\vspace{2pt}
\subsubsection{Relative Entropy Change (REC)}
\label{doa:entropy}
Entropy can be used to determine the information content of a set of values by normalizing each value by the sum of all values in the set, and treating each normalized value as its probability~\cite{palpanas2001entropy}.
We allow for a user to set bounds on minimum entropies for either a single bin or a plot.
Bin entropy can be given by $entropy(bin) = - p \times log_2(p)$, where $p = \frac{y(bin)}{\sum_{b \in bins} y(b)}$, where $bins$ represents the set of bins in a visualization. Thereby, plot entropy can be given by $\sum_{bin \in bins} entropy(bin)$. 

Continuing with our example, the bin values can be converted into probabilities as \{$0.1$, $0.2$, $0.3$, $0.4$\}, and further into entropies as \{$0.33$, $0.46$, $0.52$, $0.52$\}.
The plot entropy will be $1.83$.

Entropy is a monotonically increasing metric, i.e. splitting a bin into smaller non-trivial sub-bins causes the resulting entropy to increase.
We define \emph{REC} by
$\frac{entropy(sub-bins) - entropy(bins)}{entropy(bins)}$.
 \emph{REC} is bounded from below by $0$.
A higher value indicates that the refined bins were similar to each other -- 
whereas a value closer to $0$ indicates that the bins were dissimilar to each other and therefore, performing this refinement was beneficial to the user.
This metric can be applied at either plot or bin level.

In our example, entropies of combined sub-bins are $0.42$, $0.66$, $0.79$, and $0.91$, respectively, with the entropy of the refined plot being $2.78$. 
The plot \emph{REC} is $\frac{2.78 - 1.83}{1.83} = 0.52$. The \emph{REC} of individual bins would be $\frac{0.42 - 0.33}{0.33} = 0.27$, $\frac{0.66 - 0.46}{0.46} = 0.43$, $\frac{0.79 - 0.52}{0.52} = 0.52$, and $\frac{0.91 - 0.52}{0.52} = 0.75$.





\subsection{Result Ranking}
\label{doa:result-ranking}
Once a user sets the refinement grammar knobs, \project{} is tasked with the following naturally arising questions:

\begin{itemize}
  \item Which result bins should be refined?
  \item What should their refinement level be? 
\end{itemize}

To answer the first question, it is clear that the bins whose refinement results in greater information gain should be preferred. However, it is not possible to know this without actually refining the bins till the underlying dataset. Hence, we use our novel \emph{IGP} metric~(Section~\ref{doa:igp}), which is based on our \emph{MEI} metric~(Section~\ref{doa:mei}), to estimate the information gain potential of a bin.

To answer the second question, in keeping with our underlying principle of \emph{progressive refinement}, we provide results by progressively increasing the refinement level. Further refinement of a bin is stopped when doing so would violate the knobs set by the user, as elaborated by the result ranking algorithm~(Section~\ref{doa:algo}). 

\vspace{2pt}
\subsubsection{Maximum Entropy Increase~(MEI)}
\label{doa:mei}
\emph{MEI} is an entropy-based metric that measures the additional information that can be gained by refining a set of bins using the underlying dataset. 
Entropy of sub-bins will be minimized when a bin results in a single non-trivial sub-bin with identical measure value.
It will be maximized when the sub-bins are identical.
Thus, we can estimate the maximum possible entropy of a refined plot by
\begingroup
\setlength{\thinmuskip}{0mu}
\setlength{\medmuskip}{0mu}
\begin{flalign}
&-\sum_{i \in bins} \sum_{sb \in sub-bins(i)} \left( \frac{p_i}{|sub-bins(i)|} \times log\left(\frac{p_i}{|sub-bins(i)|}\right) \right)& \nonumber\\
&= -\sum_{i \in bins} |sub-bins(i)| \times \frac{p_i}{|sub-bins(i)|} \times log\left(\frac{p_i}{|sub-bins(i)|}\right)& \nonumber\\
&= -\sum_{i \in bins} p_i \times log\left(\frac{p_i}{|sub-bins(i)|}\right)& \nonumber\\
&MEI = -\sum_{i \in bins} p_i \times log\left(\frac{p_i}{|sub-bins(i)|}\right) - \left(-\sum_{i \in bins} p_i \times log(p_i) \right) & \nonumber\\
& = \sum_{i \in bins} ( p_i \times log(p_i) - p_i \times log(p_i) + p_i \times log(|sub-bins(i)|) )& \nonumber\\
& = \sum_{i \in bins} p_i \times log(|sub-bins(i)|)&\nonumber
\end{flalign}
\endgroup

Thus, \emph{MEI} possesses the important property of not needing to know the sub-bins' values -- it depends on the number of sub-bins, which is known apriori. 
\emph{MEI} forms the primary building block for \emph{IGP} as shown below.

To illustrate \emph{MEI}, let the aforementioned 4 bins in Section~\ref{doa:entropy} have have a domain size~(difference between upper and lower end points of a bin) of $100$ each. The \emph{MEI} as a result of refining all of them will be 
$ 0.1 \times log_2(100) + 0.2 \times log_2(100) + 0.3 \times log_2(100) + 0.4 \times log_2(100) = 6.64$.

\vspace{2pt}
\subsubsection{Information Gain Potential~(IGP)}
\label{doa:igp}
Since \emph{MEI} values can vary greatly between plots depending on their domain size, we define a new metric, \emph{IGP}, for contextualizing the values.
We define \emph{IGP} as the ratio of \emph{MEI} and entropy of the non-refined plot. Thus, the \emph{IGP} for a plot can be given by $\frac{\sum_{i} p_i \times log(|sub-bins(i)|)}{-\sum_{i} p_i \times log(p_i)}$. 
\emph{IGP} can be given for a bin $i$ by $\frac{log(sub-bins(i))}{-log(p_i)}$.
Bins with lower \emph{IGP} values are given greater importance by the ranking function.
As mentioned before, we use this metric in our ranking function to determine whether to show the refined bins to the user.

In our example, \emph{IGP} for the plot will be $\frac{6.64}{1.83} = 3.63$.
\emph{IGP} for our bins will be $\frac{0.1 \times 6.64}{0.33} = 2.01$, $\frac{0.2 \times 6.64}{0.46} = 2.89$, $\frac{0.3 \times 6.64}{0.52} = 3.83$, and $\frac{0.4 \times 6.64}{0.52} = 5.11$, respectively.
 
\vspace{2pt}
\subsubsection{Result Ranking Algorithm}
\label{doa:algo}
In our system, the cost of applying a filter to a binned dataset is much greater than that of performing aggregation.
This results in a filter query resulting possibly in more sub-bins than specified by the \emph{MaxNR} constraint.
Hence, we might need to select a subset of sub-bins to display.
We approach this problem by ranking the bins using \emph{AD} and \emph{IGP}, and displaying the top \emph{MaxNR} bins.

Note that \emph{AD} represents the benefit of refining a bin into the current sub-bins, whereas \emph{IGP} \emph{estimates} the benefit of refining the sub-bins till the highest resolution~(original non-binned data).
However, both these metrics cannot determine the true information gain possible~(refining a bin till the highest resolution).

While it is possible to use either of these metrics to rank the results, we use a commonly-used heuristic of averaging the ranks as a result of using each individually~\cite{fishburn1974sum, siegel1960nonparametric, festinger1946significance}, as we found this approach to perform the best~(Section~\ref{expt:ranking:effectiveness}).
The highest ranked results are then displayed.




\subsection{Generalized Refinement Operator~(GRO)}
\label{doa:auto-zooming}
We have seen that a user can indirectly choose the results to display by setting thresholds for the \emph{refinement grammar} knobs. 
While we would like to combine these knobs in a conjunctive fashion
to ensure that none of them are violated, this might not always be possible. 
For example, suppose that \emph{MinRef} results in 100 bins, while \emph{MaxNR} is set to 50.
Clearly, it is not possible to satisfy both constraints.

To solve this problem, we use a simple approach following Occam's Razor.
We rank the knobs based on their intuitiveness to a user and make sure that a more intuitive knob is not violated by a lesser intuitive one.
Knobs are ranked in the following order -- refinement levels, number of results, \emph{AD}, and \emph{REC}.

\subsubsection{GRO Algorithm} If \emph{MaxNR} lies between the number of results obtainable at \emph{MinRef} and \emph{MaxRef}, we use \emph{AD} and then \emph{REC} to determine the bins to refine further. 
If the current number of bins is larger than \emph{MaxNR}, we use the aforementioned \emph{result ranking algorithm}.
If \emph{MaxNR} is not specified, we refine results till \emph{MinRef} is satisfied.
They are further refined till \emph{MaxRef} if \emph{AD} and \emph{REC} are not violated.
If refinement levels are not specified, we refine results till \emph{AD} and \emph{REC} are not violated.
Thus, we can see that \emph{GRO} is the culmination of all the techniques described so far.

\subsection{Useful Refinement Operators}
\label{handy-operators}
In addition to \emph{GRO}, we also provide simple, single-click operators that serve different purposes.

\subsubsection{Refine till Highest Resolution}
This operator refines results till the non-binned dataset is queried. It follows the progressive refinement principle and provides results over intermediate refinement levels along the way.

\subsubsection{Run on Highest Resolution}
This operator simply runs the query on the non-binned dataset. 
It allows a user to opt out of progressive refinement.

\subsubsection{Refine until Non-interesting Results}
This operator stops refining a bin when it results in non-interesting sub-bins. 
We use \emph{AD} as our interestingness measure, with the interestingness threshold set to $0.1$ in accordance with Weber's law for detecting visually interesting results~\cite{weber1996eh, legge1981power}. 


 





\subsection{Parallels with Online Aggregation}
Progressive refinement provides results having smaller bin sizes over time, without any result error. Of course, there is no free lunch -- the uncertainty is encapsulated in the bin sizes. 
On the other hand, online aggregation provides results with sampling error by using a sample of the underlying dataset.
By processing more data over time, the errors usually decrease. 
Thus, we can draw parallels between these approaches -- they both result in \emph{errors}, which decrease over time. 
The errors are over the \emph{x-axis} in the progressive refinement case, while they are over the \emph{y-axis} in the online aggregation case.

\subsection{Pros \& Cons of Progressive Refinement}
While progressive refinement presents a novel, powerful approach, it is not without its pitfalls. We summarize both its benefits and shortcomings below.

\subsubsection{Pros}
Progressive refinement generally reduces the computational load on the system~(Sections~\ref{expt:computation-time}).
It also decreases the cognitive load on the user, which we define using the number of results displayed to answer a user's filter and corresponding refine queries~(Sections~\ref{expt:number-of-results} and~\ref{expt:misc:refinement}).
Its response time is low, and depends on the size of the lowest resolution dataset~(Sections \ref{expt:computation-time} and \ref{expt:user:time}).
An important consequence of a data binning-based approach is that the size of the underlying data~(number of rows) does not greatly affect the size of the binned datasets -- they are more affected by the domain size and the bin resolution.

\subsubsection{Cons}
One of the downsides is the offline pre-processing needed to compute the binned datasets. 
Determining the bins is also not straightforward as different filters are helped by different bin boundaries -- a query will be answered precisely only when the filters align with the bins. 
A binning-based approach is also applicable only for algebraic and distributive measures~\cite{malinowski2008advanced}.

\section{System Architecture}
\label{sys_arch}
\project{} uses the standard client-server architecture, and comprises of 3 layers -- \emph{frontend} to query the data and view the results, \emph{middleware} to translate user queries into progressive refinement queries that can be run on the \emph{backend}, and \emph{backend} to run the queries.
While we could use any of the previously built binning-based systems,
we use  Crossfilter~\cite{crossfilter}, as it well-suited to our session-based querying use case, and does not suffer from the cube size explosion problem.
While crossfilter has low response times for session-based queries, it has a limitation of being single-threaded and therefore cannot take advantage of the modern multi-core processing power -- we rectify this issue through the standard technique of horizontal data sharding and parallelization~\cite{green2012web}. 

\begin{figure}[h!]
	\centering
   \captionsetup{justification=centering}
   \includegraphics[width=\columnwidth]{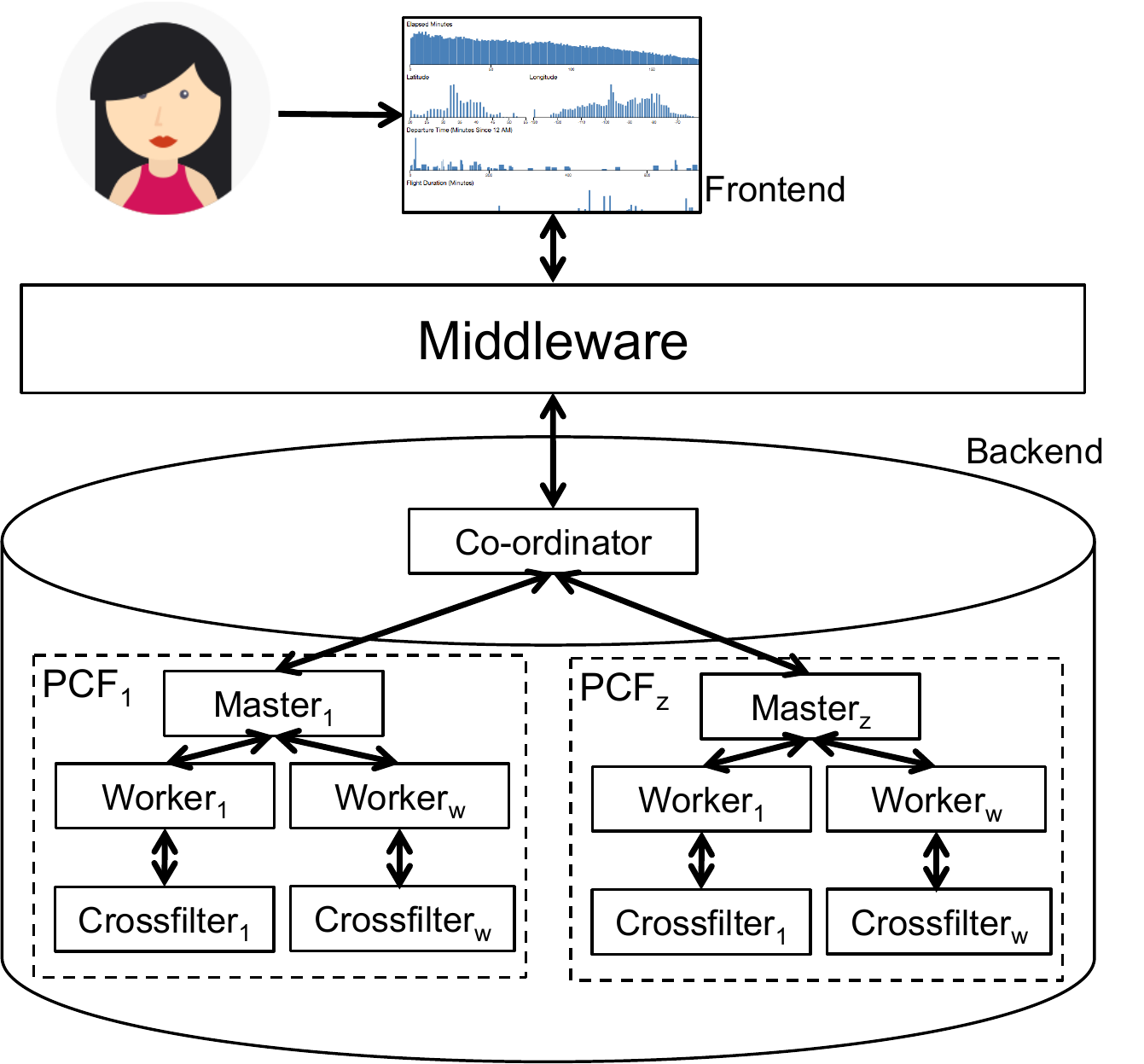}
   \caption{\project{} System Architecture consists of 3 layers -- Frontend, Middleware, and Backend -- and employs a novel parallelized crossfilter at the Backend.}
   \vspace{-10pt}
   \label{fig:avgerror}
\end{figure}

\subsection{System Components}
We now describe the various components of \project{} in more detail.

\subsubsection{Frontend} 
A user interacts with \project{} through the frontend. She can issue different filter and refine queries, which are passed on to the middleware, which queries the backend and returns the results to the frontend.

\subsubsection{Middleware}
The middleware, which runs on a \emph{Node.js} server~\cite{tilkov2010node}, interprets the user action, determines the queries that must be run, and dispatches them to the backend. Upon receiving the results back, it determines the results that must be displayed to the user, and dispatches them to the frontend.

\subsubsection{Backend}
In an offline pre-processing step, the dataset is binned into multiple smaller datasets, which are then sharded horizontally.
A parallelized multi-process crossfilter instance is created for every binned dataset, with each process running a crossfilter on its allocated shard.

At run-time, the backend end-point, termed \emph{Co-ordinator}, receives a query from the middleware, consisting of a set of filters and the resolution of the binned dataset on which they must be applied.
\emph{Co-ordinator} passes on the query to the \emph{Master} process of the specified parallelized crossfilter. Upon receiving the results from the \emph{Master}, it passes on the results to the middleware.


\noindent \textbf{Master:} 
\emph{Master} passes the query to its workers. Once it receives results from all workers, it aggregates the results,  and returns the combined result to the \emph{Co-ordinator}.

\noindent \textbf{Workers:}
Upon receiving the filters from its \emph{Master}, each \emph{Worker} applies them to its crossfilter and returns the results.

\subsection{Binning Strategies}
While the concept of binning using histograms has a rich history~\cite{pearson1894contributions, howitt2007introduction}, determining the ideal binning strategy has remained an elusive problem. 
Different strategies have different benefits -- wider bins reduce noise for low density areas at the cost of lower precision, while narrower bins provide higher precision for high density areas while increasing the effect of noise. 

These considerations for static visualizations are further complicated in the dynamic case.
In this context, our progressive refinement concept, which varies bin definitions over time, thereby providing the results at varying resolutions according to the user's requests, can be seen as an attempt at solving the bin determination problem.




There exist two general strategies for binning -- \emph{equi-width} and \emph{equi-data}.
In the \emph{equi-width} case, the domain is divided equally between bins, whereas in the \emph{equi-data} case, each bin consists of approximately equal number of tuples.
Our preliminary user studies guided us towards using \emph{equi-width} binning since it results in the changes to x-axis being smoother and more intuitive, thereby allowing a user to focus on the changes in the y-axis.
On the other hand, in the \emph{equi-data} case, changes occur to the bin ranges on the x-axis, which are more difficult to visually analyze.

Currently, bins for different dimensions are determined independently of each other as it is unclear whether the added complexity is worthwhile -- we use the marginal distribution along a dimension in determining its bins.
In the future, we would like to
consider prior workloads in determining the bins.

\section{Experiments}
\label{expt}
We evaluated \project{} extensively through real and simulated workloads over real and benchmark datasets, using numerous metrics, some which are the execution time, number of results displayed, and the number of queries executed and hypotheses tested.

\subsection{Experimental Setup}
Users interact with \project{} through its user interface on the client machine -- 
a Ubuntu Linux 16.04.3 LTS system 
with a 4-core 3.3GHz Intel Core CPU, 
16GB DDR3 @ 1600 MHz memory, 
and a 256GB @ 7200 RPM disk.
The datasets, as given in Table~\ref{table:datasets}, are loaded in our parallelized version of Crossfilter 1.3.12 running on Node.js 7.4.0 on our server -- a Ubuntu Linux 14.04.1 LTS system with a 24-core 2.4GHz Intel Xeon CPU, 256GB DDR3 @ 1866 MHz memory, and a 500GB @ 7200 RPM disk, which communicates over a 1 Gbps network with the client. 

\subsection{Datasets}
\label{expt:datasets}
We evaluated \project{} using 5 datasets as given in Table~\ref{table:datasets}, with 3 of them being real-world datasets -- a private aviation dataset~(\emph{PAD}), \emph{Flights}~\cite{flights}, and \emph{Brightkite}~\cite{cho2011friendship}. 
\emph{SPLOM}~\cite{kandel2012profiler}, the standard benchmark in interactive data cubing, was used to generate two datasets having 10M and 1B rows each.
To maintain uniformity across datasets in our experiments, each dataset was used to create 5 binned datasets~(refinement levels from 0 to 4), with each split generating 2 sub-bins from a bin.
As users can query the underlying non-binned dataset as well, this results in a total of 6 refinement levels for each dataset.





\subsection{User Study Setup}
\label{expt:user-study-setup}
We designed our user study to understand user behavior in exploration of large datasets through the progressive refinement paradigm, and evaluate the benefits and short-comings of \project{}.
Users were asked to explore the \emph{PAD} dataset and extract possible insights.
They were also asked to report any hypotheses that they might be testing, and whether their hypotheses were validated or invalidated by their queries.
The participants consisted of 12 graduate students pursuing their PhD.
All participants, except 2, were conducting research in the fields of either databases or data mining, and thus had a background in data analysis.



Each user study consisted of two sessions.
In one of the sessions, users were asked to explore the dataset using the full-fledged \project{} system using filter and refine queries. 
In the other session, as part of the base case, users analyzed the underlying non-binned dataset using only filter queries, without any of the progressive refinement features. 
To control for learning and order effects, session order was randomized.
Each experiment lasted for a minimum of 5 minutes. 
Users could continue exploration at the end of the 5 minutes, if they chose to.
There was a 5 minute break between the two sessions.
All user actions performed during the study were logged.
At the end of the sessions, multiple metrics were computed.


\subsection{Simulated Querying Setup}
\label{expt:automated-querying}
In addition to the user study, to study \project{} in more detail, we simulated user behavior through queries generated in an automated fashion, resulting in 100 queries for each of the datasets.
As modeling complex user behavior is a non-trivial task, we employed a simple, generalized model,
where a user changes filters on different plots, and then increases the refinement level incrementally over all visualizations.
Each workload was executed 3 times, with the caches being flushed before every run.
The presented results are their averages over the 3 runs.



\subsection{Workloads}
The simulated queries resulted in the following four workloads -- \emph{PAD\_Auto}, \emph{Flights}, \emph{Brightkite}, and \emph{SPLOM\_10M}. 
To study the user behavior sessions in concert with the simulated sessions, we modified the user sessions as follows -- we inserted refinement queries similar to those described in Section~\ref{expt:automated-querying} after every filter query. 
Refinement queries issued by the user were removed.
This gave us the \emph{PAD\_Progressive} and \emph{PAD\_Base} workloads.
User sessions, in their non-modified form, are studied in more detail in Section~\ref{expt:user-study-results}.

\begin{table}
\hfill{}
\begin{tabular}{|c | c | c | c |} 
\hline
Dataset & Refinement & \multirow{2}{*}{$\vert \textrm{Rows} \vert$} & File\\
(\# Dimensions) & Level & & Size\\ 
\hline
\hline
\multirow{6}{*}{PAD~(8)}                           & 0            & 64K                        & 2M\\ 
                              & 1            & 1M                     & 37M\\ 
                              & 2            & 10M                    & 332M\\ 
                              & 3            & 28M                    & 865M\\ 
                              & 4            & 38M                    & 1.2G\\
                              & base         & 50M                    & 1.5G\\
\hdashline 
\multirow{6}{*}{Flights~(6)}        & 0           & 755                      & 22K\\ 
                              & 1           & 21K                      & 583K\\ 
                              & 2           & 869K                     & 23M\\
                              & 3           & 17M                      & 434M\\
                              & 4           & 81M                      & 2.0G\\
                              & base        & 121M                     & 2.5G\\
\hdashline
\multirow{6}{*}{Brightkite~(4)}        & 0           & 336            & 7K\\ 
                              & 1           & 22K                      & 425K\\ 
                              & 2           & 574K                      & 11M\\
                              & 3           & 3M                      & 56M\\
                              & 4           & 4M                      & 80M\\
                              & base    & 4.7M                      & 86M\\
\hdashline
\multirow{6}{*}{SPLOM\_10M~(5)}        & 0           & 1018              & 28K\\ 
                              & 1           & 10K                      & 275K\\ 
                              & 2           & 102K                      & 3M\\
                              & 3           & 841K                      & 21M\\
                              & 4           & 4M                      & 102M\\
                              & base    & 10M                      & 235M\\
\hdashline
\multirow{6}{*}{SPLOM\_1B~(5)}         & 0           & 1.5K             & 43K \\ 
                & 1           & 19K             & 525K \\ 
                & 2           & 226K             & 6M \\ 
                & 3           & 2.46M             & 63M \\ 
                & 4           & 22.78M             & 566M \\ 
                & base\footnotemark        & 1B             & 22G \\ 
\hline
\end{tabular}
\hfill{}
\caption{Datasets.}
\vspace{-20pt}
\label{table:datasets}
\end{table}

\footnotetext{
Querying individual tuples of the underlying SPLOM\_1B dataset is currently not possible in \project{} due to the memory requirements of crossfilter. 
Hence, the results for \emph{SPLOM\_1B} are not provided -- while it is possible to query the binned datasets, the baseline results are unavailable.}

\subsection{Results}
\label{expt:auto-metrics}
We evaluate the benefits provided by the progressive refinement paradigm over the base case~(querying the underlying dataset) exhaustively using multiple, complementary metrics.


\subsubsection{Reduction in Computation Time~(RCT)}
\label{expt:computation-time}
\begin{figure}[h!]
   \centering
   \captionsetup{justification=centering}
   \includegraphics[width=\columnwidth]{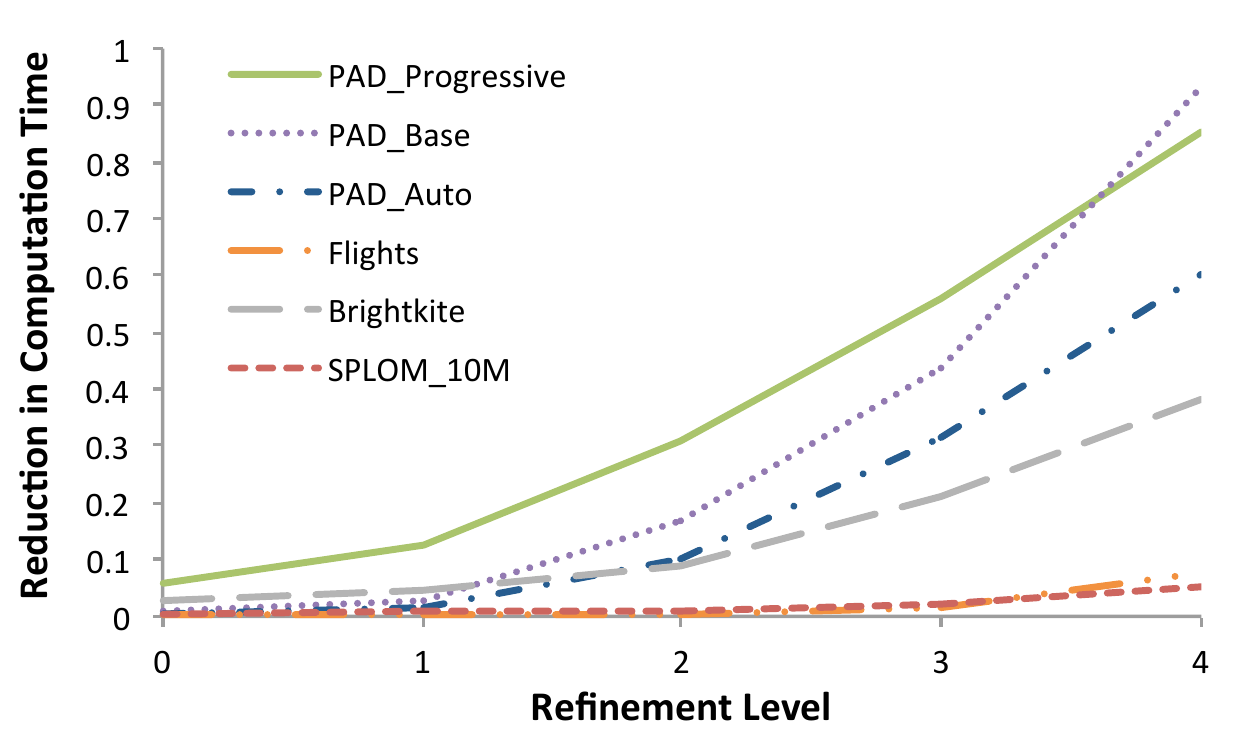}
   \caption{Computation Time.}
   \label{fig:rct}
\end{figure}

The biggest benefit that progressive refinement provides is the reduction of the execution time as queries do not need to hit the underlying dataset.
We define \emph{RCT} as the ratio of the cumulative time taken to answer a query through the progressive refinement paradigm, to the time taken by the query running over the underlying data.
A lower value indicates that the user query was satisfactorily answered at a lower computational cost, while a value larger than 1 indicates that progressive refinement might not have been useful.
Figure~\ref{fig:rct} shows that while the cumulative execution time increases with the refinement level, it is still lower than the time taken to run a query over the non-binned dataset.
Further, the initial response time is extremely low for all workloads.

\begin{figure}[h!]
   \centering
   \captionsetup{justification=centering}
   \includegraphics[width=\columnwidth]{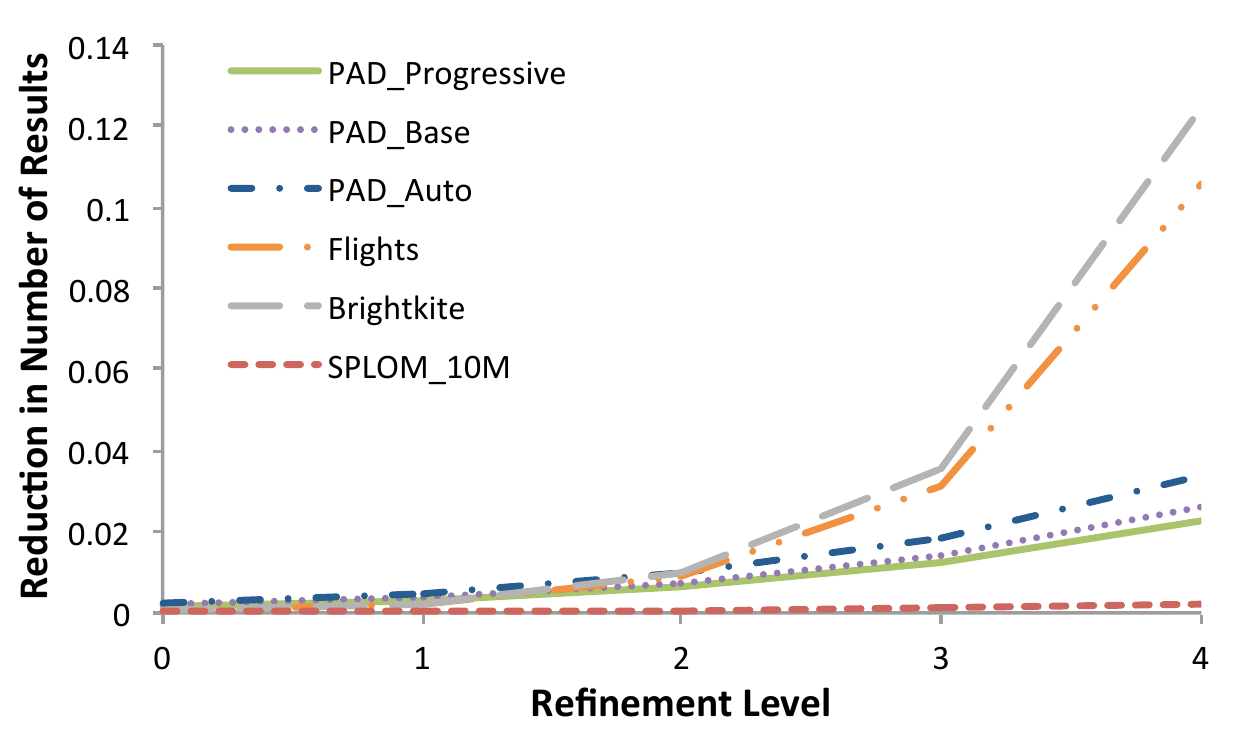}
   \caption{Number of Results.}
   \label{fig:rnr}
\end{figure}

\subsubsection{Reduction in Number of Results~(RNR)}
\label{expt:number-of-results}
By reducing the number of irrelevant results, \project{} reduces the cognitive load on the user, allowing her to focus on the more interesting results.
We define \emph{RNR} as the ratio of number of results shown to the user to the number of results that can be obtained by running the query over the underlying dataset. 
Figure~\ref{fig:rnr} shows that \emph{RNR} is low -- even at the refinement level of 4, \emph{RNR} is at least an order of magnitude smaller than 1 for all workloads.


\subsubsection{Result Error~(RE)} 
\label{metric:result_error}
\begin{figure}[h!]
   \centering
   \captionsetup{justification=centering}
   \includegraphics[width=\columnwidth]{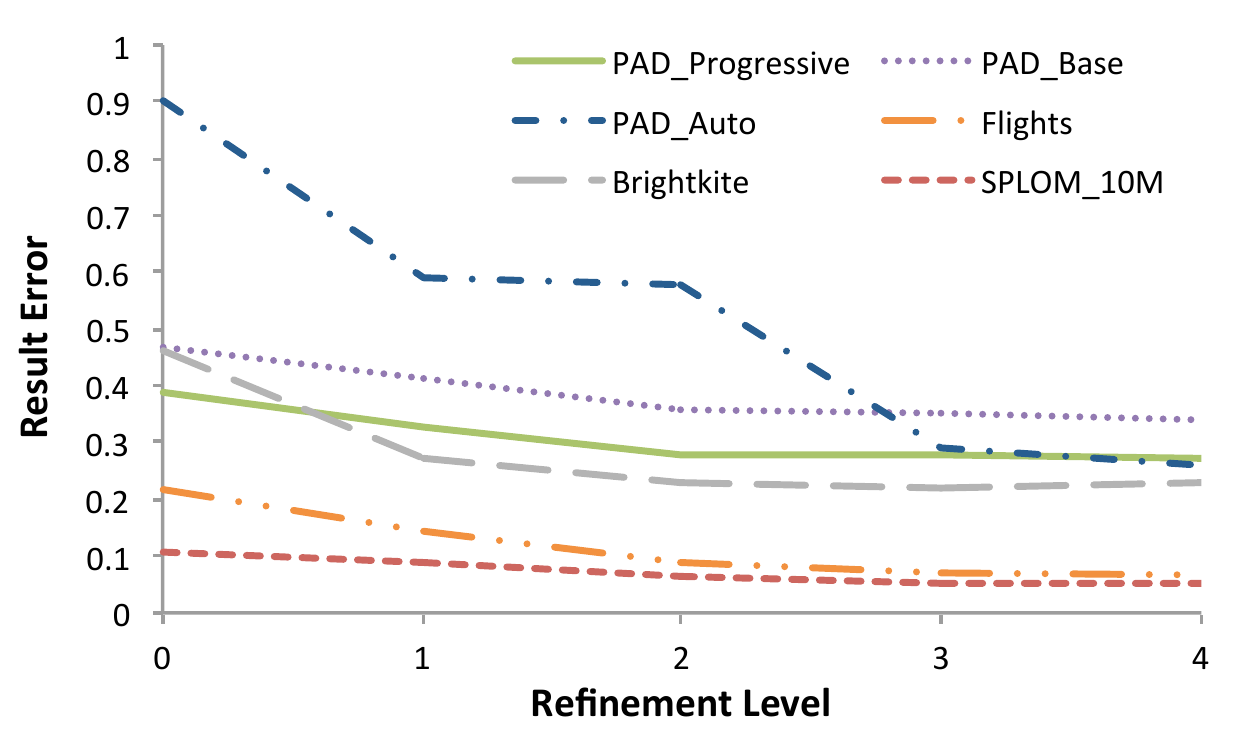}
   \caption{Result Error.}
   \label{fig:re}
\end{figure}

A binned result can be used to estimate its refined results
using uniform distribution~(Section~\ref{doa:AD}).
The true value of refined results can be determined by running the query over the underlying non-binned dataset.
While the binned results are themselves accurate, \emph{RE} captures how well the bins reflect the results over the underlying data.
We define \emph{RE} for a bin by $Average_{sb \in sub-bins}|\frac{sb - esb}{esb}|$, where \emph{sub-bins} represents the results that lie within the bin that are obtained by running the query over the non-binned dataset.
$esb$ represents the expected refined result under the uniform distribution assumption.
Figure~\ref{fig:re} shows that \emph{RE} generally decreases over increasing refinement levels with low enough errors even at the level of 1 for some workloads. 

\subsubsection{Anomalous Fraction~(AF)}
\label{expt:anomalous}

\begin{figure}[h!]
	\centering
   \captionsetup{justification=centering}
   \includegraphics[width=\columnwidth]{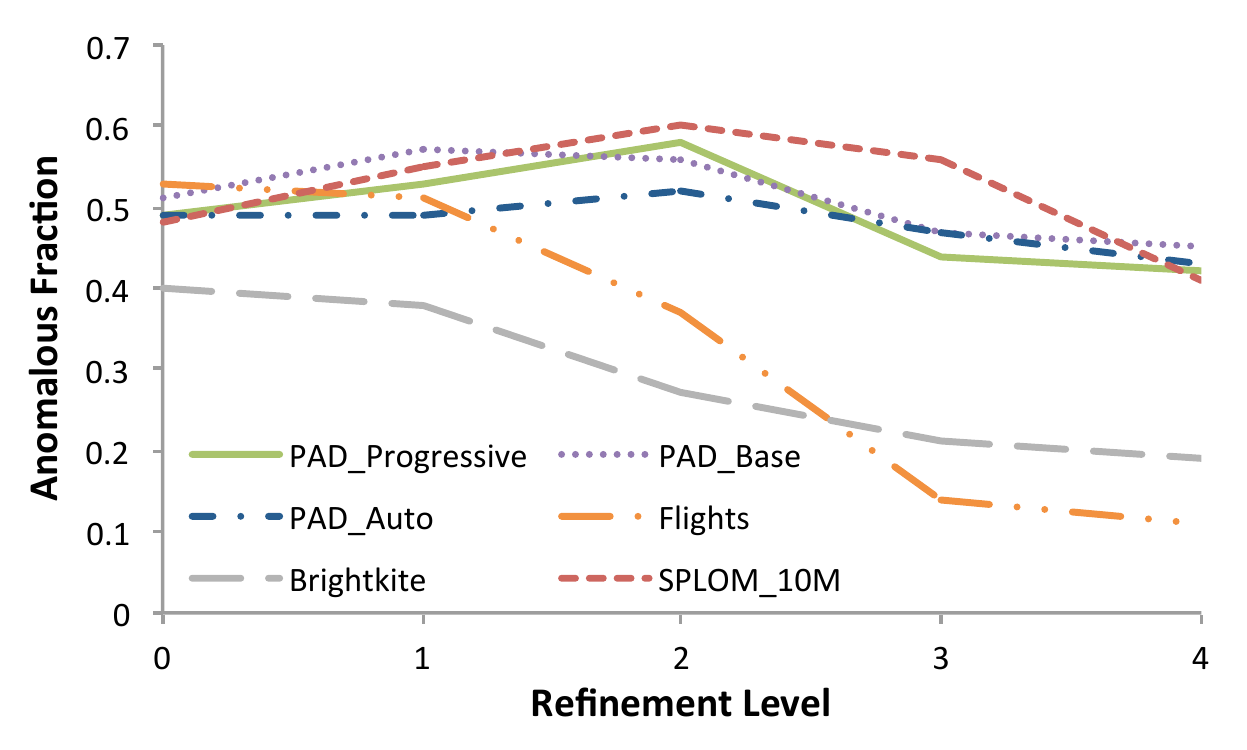}
   \caption{Anomalous Results.}
   \label{fig:anomalous-results}
\end{figure}
Analyzing anomalous results is an important part of data analysis.
While there exist different complex techniques to determine anomalous results such as using Lorenz curve~\cite{mcgarry2005survey}, p-value~\cite{wasserstein2016asa}, Gini co-efficient~\cite{gini1997concentration}, etc. we use a simple context-dependent metric -- we term a result as being anomalous if it is significantly different from its neighbors, i.e. $Anomalous\ results = \{ x : \forall n \in neighbors(x), \  |\frac{n - x}{x}|> 0.1 \}$. 
Figure~\ref{fig:anomalous-results} shows that in most cases, the number of anomalous results decreases with increasing refinement levels -- the number of results increases with increasing refinement levels, while the number of underlying anomalous results stays constant. 
We cannot explain increase in this metric at refinement levels 1 and 2 for some workloads.

\subsubsection{Effectiveness of Ranking Techniques}
\label{expt:ranking:effectiveness}

\begin{figure}[h!]
   \centering
   \captionsetup{justification=centering}
   \includegraphics[width=\columnwidth]{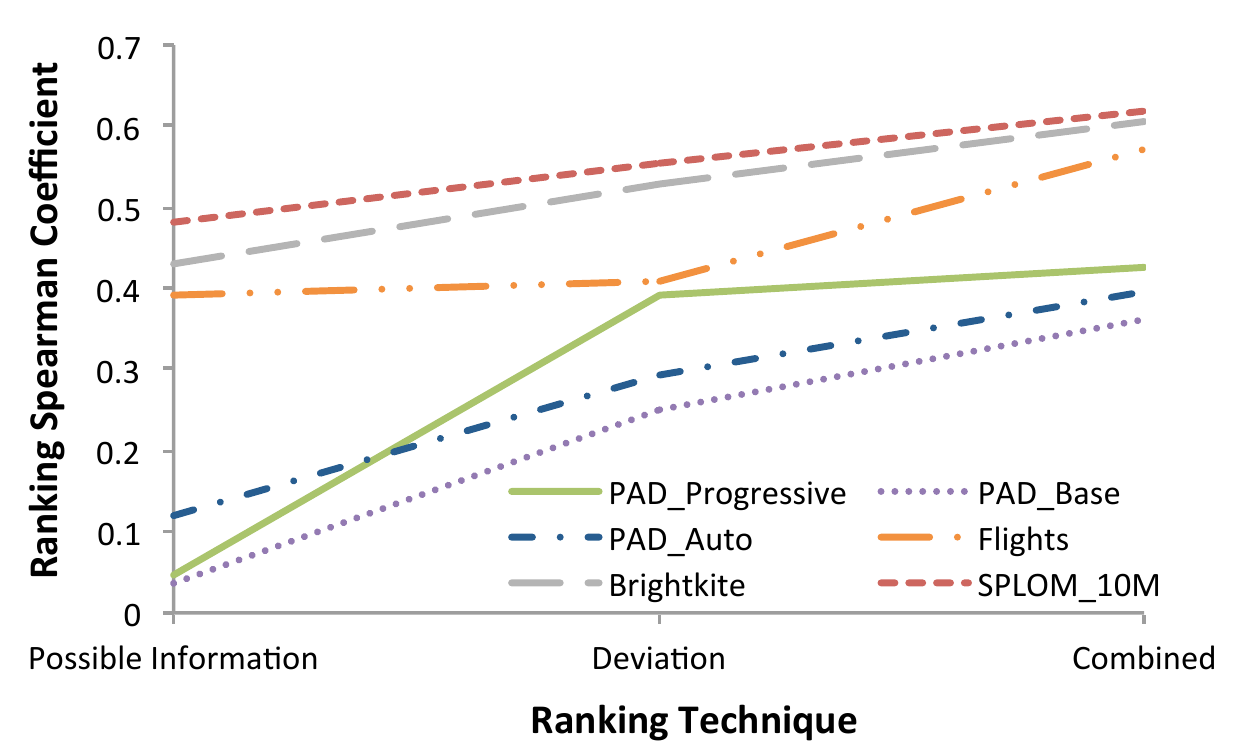}
   \caption{Ranking Effectiveness.}
   \label{fig:spearman}
\end{figure}

As a user can limit the number of results to display, \project uses a novel result ranking technique as detailed in Section~\ref{doa:algo}.
The true importance of a result is determined by it's \emph{RE} using the underlying dataset-- those with larger \emph{RE}s are given greater importance.
We present \emph{Spearman's Correlation Coefficient}~\cite{myers2010research} for each of our ranking techniques, which as described in Section~\ref{doa:algo}, consist of using only \emph{AD}, or only \emph{IGP}, or their average rank.
Interestingly, Figure~\ref{fig:spearman} shows that using the average rank results in a better ranking scheme than using either of the techniques individually for all workloads.

\subsubsection{Relative Entropy Change} 
\begin{figure}[h!]
   \centering
   \captionsetup{justification=centering}
   \includegraphics[width=\columnwidth]{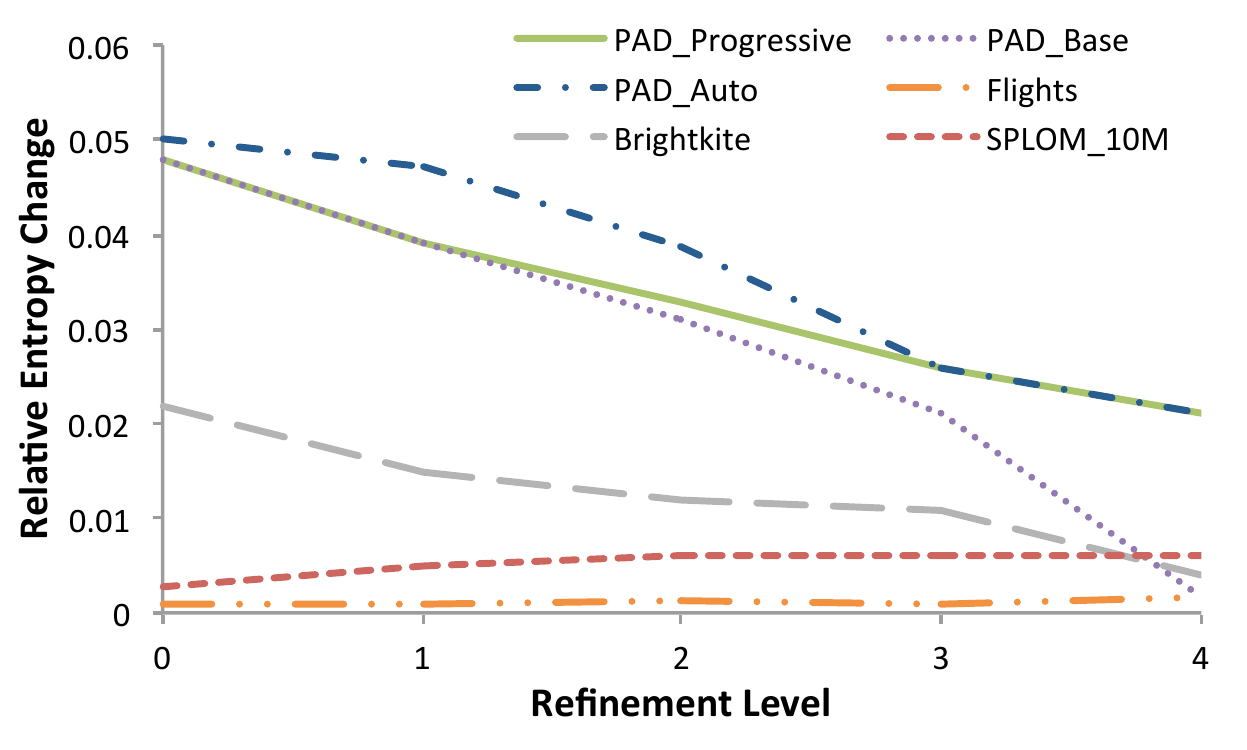}
   \caption{Relative Entropy Change.}
   \label{fig:rec}
\end{figure}

We determine the average information loss as a result of not refining till the highest level using the average of \emph{REC}s~(Section~\ref{doa:entropy}).
Figure~\ref{fig:rec} shows that this metric generally decreases with increasing refinement levels -- this is due to the results at higher refinement levels having greater resemblance with the result over underlying dataset, causing entropies to be similar.

\subsubsection{Data Sparsity}
\begin{figure}[h!]
   \centering
   \captionsetup{justification=centering}
   \includegraphics[width=\columnwidth]{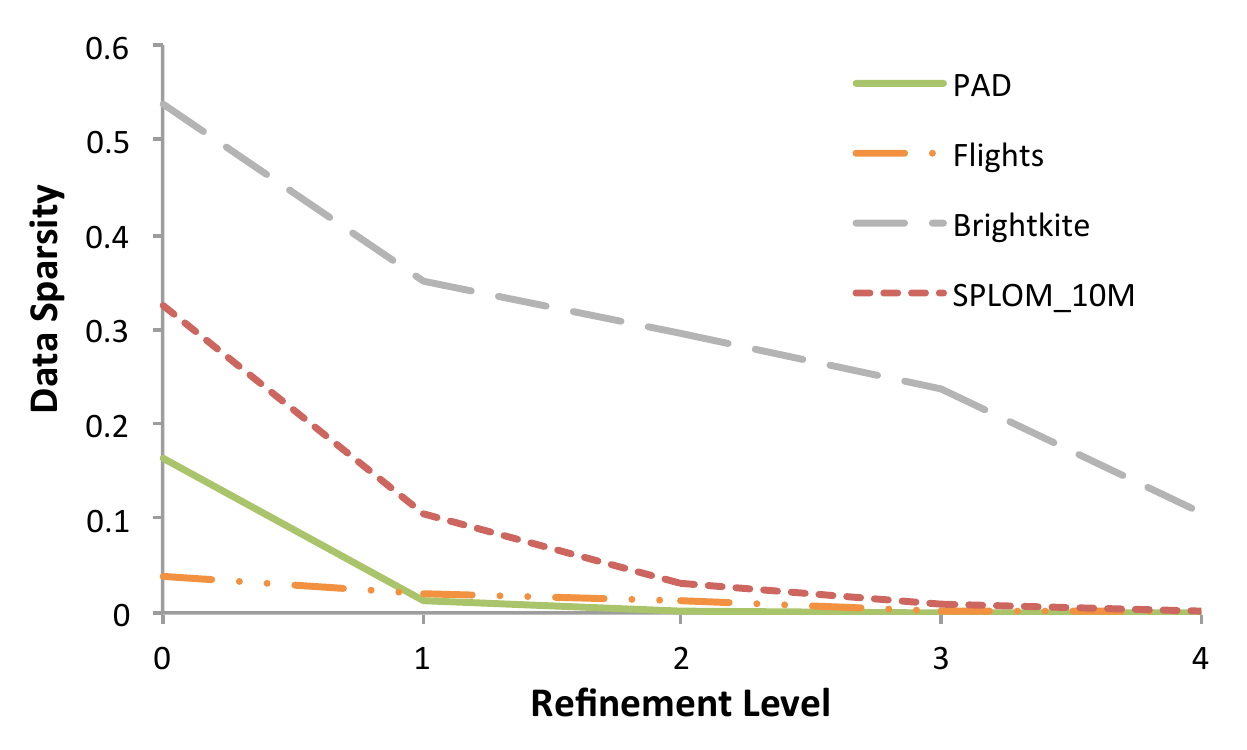}
   \caption{Data Sparsity}
   \label{fig:sparsity}
\end{figure}

Since data binning forms an integral part of this paper, we look at data sparsity, i.e. the ratio of the number of rows in a binned dataset to the maximum number of rows possible, which given cardinality $d_i$ for the $i^{th}$ dimension can be given by $max\_rows = \prod_{i = 1} ^ {d} d_i$, for $d$ dimensions. 
Unless there exist $max\_rows$ distinct tuples, some bins can be expected to be empty. 
Due to the curse of dimensionality, we would expect this ratio to decrease with increasing resolution levels, which Figure~\ref{fig:sparsity} indeed demonstrates.






\subsection{User Study-Specific Results}
\label{expt:user-study-results}
In this section, we analyze the user study results in a detailed fashion. 
Note that the sessions consist of user-specified filter and refine queries.
We measure multiple metrics for every user query session, and aggregate them over sessions through their average and median.
Table~\ref{table:misc-userstudy} summarily demonstrates how the progressive refinement paradigm improves upon the base case~(querying the underlying non-binned dataset).
We note that the results are statistically significant~($p = 0.05$), even for the stronger hypothesis of the metric in the progressive refinement case being greater, or appropriately lesser, than the base case, for all metrics except \emph{Session Duration}.
In discussing the results, we use the median value instead of the average to account for outliers, although both values are similar for most of the metrics.

\begin{table}[h!]
\hfill{}
\begin{tabular}{|c|c|c|c|c|c|c|}
\hline
\multirow{2}{*}{Metric}&\multicolumn{2}{c|}{Median}&\multicolumn{2}{c|}{Average}&\multicolumn{2}{c|}{StdDev}\\
\cline{2-7}
&Prog&Base&Prog&Base&Prog&Base\\
\hline
Avg Query Time (s) & 0.07 & 27.03 & 0.08 & 24.02 & 0.046 & 17.35 \\
\hline
Total Query Time (s) & 2.64 & 147.3 & 2.64 & 169.2 & 1.31 & 68.16 \\
\hline
Query Time Fraction & 0.004 & 0.47 & 0.004 & 0.51 & 0.004 & 0.144 \\
\hline
\# Filter Queries & 30.5 & 5 & 38.7 & 5.2 & 22.84 & 1.3 \\
\hline
Session Duration (m) & 8.5 & 5.2 & 8.5 & 5.6 & 2.7 & 0.89\\
\hline
\# Hypotheses & 7 & 1 & 8.91 & 1.8 & 6.2 & 1.3 \\
\hline
\end{tabular}
\hfill{}
\caption{User Study Results.}
\label{table:misc-userstudy}
\end{table}



\subsubsection{Query Execution Time}
\label{expt:user:time}
Providing low query execution times is a central requirement for a interactive analysis system.
We can see that the progressive refinement paradigm took $0.07$s to execute a query compared with $27.03$s taken by the base case.
Thus, \project{} not only provides multiple orders of magnitude speedup over the base case, but does so within interactive response times.

This results in $47\%$ of the session time being spent running the query in the base case compared with $0.4\%$ for progressive refinement. 
Wasting such a large fraction of an analyst's time is clearly inconducive for interactivity -- as we will see, this has a snowballing effect on the number of queries a user executes in a session and consequently more importantly, the number of hypotheses she is able to investigate.

\subsubsection{Filter Queries}
We now look at the number of filter queries in each session.
Section~\ref{expt:misc:refinement} discusses refinement queries.
While $30.5$ queries could be run through progressive refinement approach, only $5.2$ queries could be run in the base case.
This shows that users were often content with the results at lower resolution as in the progressive refinement case, and as a result were able to explore multiple hypotheses quickly.

\subsubsection{Session Duration}
While users were made aware of when their session reached the 5 minute mark, most users continued exploring for a total duration of $8.5$ minutes in the progressive refinement case, in comparison with the $5$ minutes spent in the base case.
We attribute this extra time spent by busy graduate students to their curiosity in analyzing the dataset, utility of the progressive refinement paradigm, and the usefulness of \project{} in helping them do so.
Note that even after normalizing for the session duration, number filter queries issued is significantly larger in the progressive refinement case.

\subsubsection{Hypothesis Testing}
An important functionality that any data exploration system should provide is facilitation of \emph{hypothesis testing}, i.e. being able to quickly form and validate hypotheses.
As mentioned in Section~\ref{expt:user-study-setup}, users informed us of their hypothesis, which they tested through filter and refinement queries.
Users were able to test $8.9$ hypotheses through the progressive refinement paradigm, in comparison with a single hypothesis in the base case.
Note that we do not refer to the statistical sense of the term hypothesis testing.

\subsubsection{Refinement Queries}
\label{expt:misc:refinement}
In this section, we look at different refinement query statistics.
We note that only $16.4\%$ of the queries refined the data -- a large majority of the queries were filters.
It is thus imperative that an optimal initial refinement level be chosen in a progressive refinement system -- it should be detailed enough to not need further refinement, but low enough to return results within interactive times.

Refinement queries resulted in $7.18\%$ additional results being displayed. Thus, queries could be answered with fewer results, thereby, \emph{reducing users' cognitive load}.
 

They also resulted in a $7.85\%$ increase in the execution time.
The lower increase in the execution time is due to some refinement queries being only over single plots, while each filter query needing to modify all plots. 



\subsubsection{Query Complexity}
The number of filters~(\texttt{WHERE} predicates) is a good indicator of query complexity -- more filters indicates a more detailed and a more complex query. Table~\ref{table:filters} shows the percentage of queries having different number of filters. 
We can see that in the progressive refinement case, some queries had filters on up to 5 dimensions, while the base case had a maximum of 3 filters in a query -- this is to be expected as the iteration speed allows users to test more complex hypotheses. Thus, not only are the users able to issue more queries, but the queries are more complex as well.

\vspace{-6pt}
\begin{table}[h!]
\hfill{}
\begin{tabular}{| c |  c | c | c | c | c | c | c | c | c |} 
\hline
\# Filters & 0 & 1 & 2 & 3 & 4 & 5 & 6 & 7 & 8\\
\hline
\%Queries\_Prog & 19.2 & 54.1 & 10.8 & 7.8 & 4.8 & 2.9 & 0 & 0 & 0\\
\hline
\%Queries\_Base & 15.3 & 57.7 & 21.2 & 5.7 & 0 & 0 & 0 & 0 & 0\\
\hline
\end{tabular}
\hfill{}
\caption{Fraction of Queries by Number of Filters.}
\label{table:filters}
\end{table}

\subsubsection{User Comments}
While multiple objective metrics provided above capture different benefits of the progressive refinement paradigm, a subjective metric such as user comments captures the intangibles and provides a complementary, perhaps richer insight.
To start off, users unanimously chose our \emph{progressive refinement} approach over the base case. 
Some illustrative remarks were:
"I will grab my coffee real quick while this is running", 
"It's too slow to query the entire data.", 
"First query and I am already annoyed",
"while waiting", 
"I did not feel seeing so many results was useful.",
"ok there it goes. This is pretty much unworkable".




\subsubsection{User Behavior \& Future Features}
Filter queries in a brushing and linking-based system such as ours fall in one of 3 categories -- adding a filter~($24.41\%$), modifying a filter~($45.82\%$), and removing a filter~($29.77\%$).

We observed that the most recently added filter was more likely to be removed or modified.
This leads us towards an important feature that a data analysis system should provide --  user guidance, and in particular, query suggestion.

Some filters resulted in multiple visualizations changing drastically -- as a user cannot visually follow changes in multiple visualizations, it is important to be able to highlight results that have undergone significant changes.

It is also important to reduce the number of refinement queries a user needs to perform -- a user pointed out "it breaks my flow". This points towards perhaps refining \emph{interesting} results after a filter query finishes execution, by default.

A progressive refinement system is thus highly affected by having an \emph{optimal} initial refinement level -- there does not seem to be an ideal strategy for determining it.
We used the heuristic of setting an interactive response time threshold and using the largest dataset that was able to meet it for most of the filter queries.
We have set this threshold for \project{} at $0.1$s -- while the interactive response time threshold can vary with datasets and systems, research has demonstrated the need for it to be no more than $500$ ms~\cite{liu2014effects}.


\subsection{Parallel Crossfilter Speedup}

\begin{figure}[h!]
   \centering
   \captionsetup{justification=centering}
   \includegraphics[width=\columnwidth]{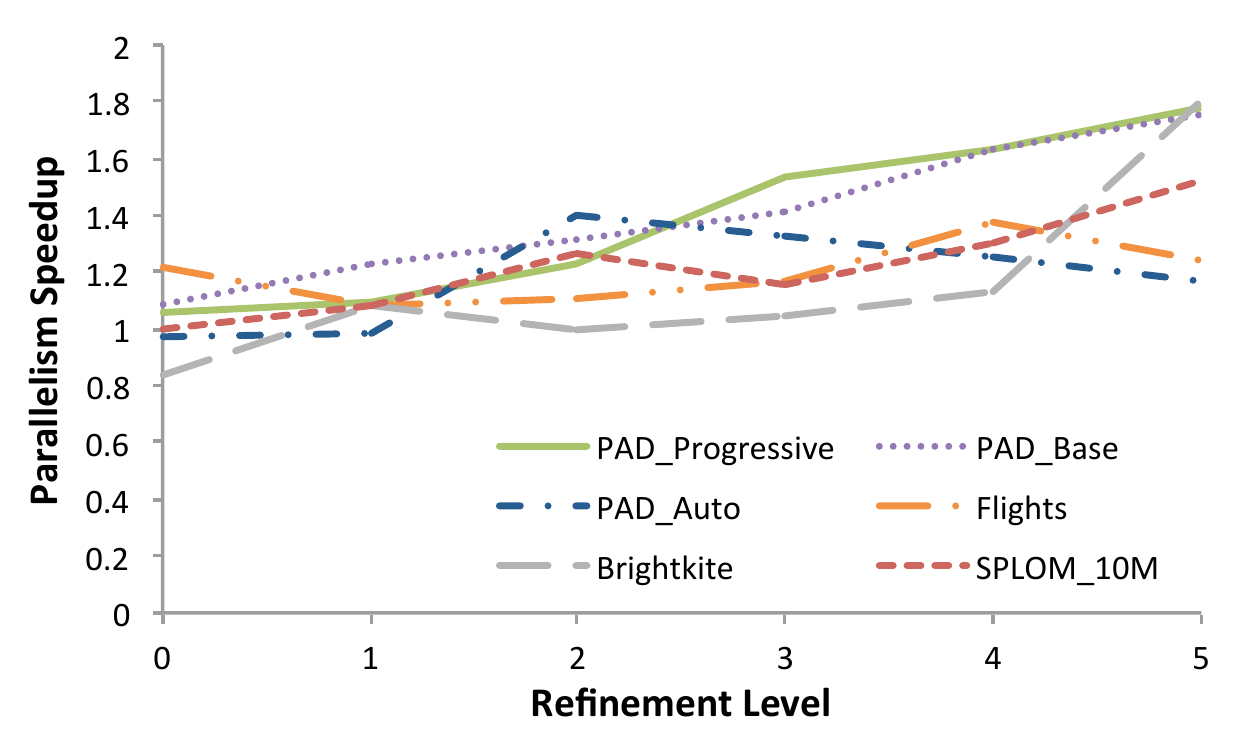}
   \caption{Parallel Crossfilter Speedup}
   \label{fig:parallel}
\end{figure}
Section~\ref{sys_arch} describes \project's system architecture -- here, we look 
at the speedup as a result of our novel parallelized crossfilter design.
We can see that parallelization can provide speedup of up to $2 \times$ for higher refinement levels -- at lower levels, due to lower execution times, the additional cost of parallelization slightly  outweighs the resultant speedup.
This guides us towards an interesting architectural design for the future -- using non-parallelized crossfilter at lower resolutions and a parallelized crossfilter for higher resolutions, with the crossover point being determined empirically.
Note that we use the standard webworkers package~\cite{webworkers} for Node.js. It does not provide linear speedup with increasing number of cores -- we profiled this to be the cause of the overall sub-linear speedup. Other packages are limited in their support for closures and are unsuitable for our purposes. In the future, we would like to develop a scalable webworkers package, which would benefit not only our parallelization framework, but more importantly, the broader web development community.

\section{Related Work}
\label{related}


\nocite{bostock2012d3}

While data cubes~\cite{gray1997data} expedite analytical queries over large datasets, their size increases exponentially with dimension cardinalities, thereby increasing the time needed for their construction and the space needed to store them.
However, more importantly from an interactive querying perspective, it affects their ability to help query large datasets within interactive response times. 

Sampling can help scale to large datasets by running queries on a representative sample of the data~\cite{olken1993random, chaudhuri1999random, agarwal2013blinkdb, kandula2016quickr}.
However, sampling introduces multiple issues in the analysis process, including sampling error, its interpretation, and visualization.
Online aggregation~\cite{hellerstein1997online, haas1999ripple,  jermaine2005disk, nirkhiwale2013sampling, li2016wander} builds upon sampling by providing results whose measure error generally decreases over time, as a result of processing more data.
Our approach is orthogonal to online aggregation -- while online aggregation decreases error over \emph{y-axis}, progressive refinement does so over the \emph{x-axis}.



Data binning for the purpose of visualizing large datasets is intuitive and has been employed by numerous systems such as Profiler~\cite{kandel2012profiler}, imMens~\cite{liu2013immens} and, \emph{Nanocubes}~\cite{lins2013nanocubes}, which construct data cubes over binned datasets.
This reduces the size of the resultant cubes.
However, due to the data cube size explosion problem, Profiler and imMens have been documented to scale upto 5 and 4 linked views, respectively.
\emph{Nanocubes}~\cite{lins2013nanocubes} allows refinement up to a fairly high-resolution version of the visualizations -- unlike \project{} however, it does not allow drilling down to the individual records.
While \emph{Nanocubes} is able to reduce the size of the data cube by the \emph{sharing} factor through smart indexing, it cannot deal with the inherent theoretical data cube size explosion problem discussed earlier. 
By avoiding building cubes over the dataset through crossfilter, \project{} sidesteps this problem -- while execution time will be low for the session-based querying scenario where subsequent queries are related to each other, it will be comparatively higher for random user queries.




While these systems provide the standard refinement-level based operator, their focus is different from ours -- Profiler helps assess quality issues in the data, imMens incorporates parallel query processing through GPUs to visualize multi-variate data tiles, while Nanocubes extends Dwarfcubes for spatiotemporal data and aims to greatly reduce the cube size -- we investigate the refinement operator in detail, and study the effect of the progressive refinement paradigm on the user.
We treat refinement as a first-class citizen and evaluate the consequent benefits of using refinement as a central operator. 
We enrich the standard refinement operator by allowing for enhancements such as limiting the number of results, and using the information content of the visualization into account in determining the results.
Our strategies represent a middleware layer -- \project{} could have used any of these systems as our backend through considerable engineering effort. 
\nocite{li2008sampling}

\emph{Hashedcubes}~\cite{pahins2017hashedcubes} provides an alternative to \emph{Nanocubes} by using a more compact representation and a simpler implementation.
\emph{Dwarfcubes}~\cite{sismanis2003hierarchical} laid foundations for compression techniques for data cubes, which \emph{Nanocubes} enriches.
Other systems such as M4~\cite{jugel2014m4, jugel2016vdda}, ScalaR~\cite{battle2013dynamic}, Forecache~\cite{battle2016dynamic}, etc. modify user queries by taking the screen resolution into consideration to not only reduce the work done at the backend, but also in transmission of the results over network.
\emph{VisReduce}~\cite{im2013visreduce} incrementally computes visualizations in a distributed environment.

Recent approaches have looked at incorporating sampling into visualizations.
\emph{VAS}~\cite{park2016visualization} provides high-quality visualizations using a small subset of the data.
\emph{Pangloss}~\cite{moritz2017trust} enumerates numerous visualization issues in approximate query processing.
Kwon et al. make the case for using sampling in visualizations and detail numerous issues important to sampling-based visual analytics~\cite{kwon2017sampling}.
\emph{ProgessiVis} enables changes at the language and library level to support exploratory analysis systems~\cite{fekete2016progressive}.
\emph{IncVisage}~\cite{rahman2017ve} builds on their \emph{zenvisage} system~\cite{siddiqui2016effortless}, by using sampling to quickly reveal salient features of a visualization, while minimizing the result error. 
Unlike \project{}, none of these systems investigate result refinement through binning, with sampling being their preferred tool for building interactive visualization systems.

\section{Conclusion \& Future Work}
\label{conclusion}
\project{} provides novel techniques for interactive visualization-driven analysis of large datasets, through our novel querying approach of progressive refinement, where we treat result refinement as a first-class operator.
Our visualizations consequently are devoid of any error in the measure, with information loss being contained in the bin sizes. 
Progressive refinement has been highly influenced by the \emph{Progressive Transmission} technique -- transmitting successively refined images to users using Image Pyramids -- prevalent in the early days of the internet with low network speeds~\cite{adelson1984pyramid}.
Progressive refinement also draws parallels with online aggregation, as it reduces error over time over the \emph{x-axis} as opposed to the \emph{y-axis}, and thus presents a new paradigm in approximate querying.
We demonstrate theoretically as well as empirically the fact that both computational load over the system and cognitive load over the user are reduced -- this results in the user being able to execute an order of magnitude more queries. The queries also posses more complexity.
This culminates in the users being able to test a significantly higher number of hypotheses. 

In the future, we would like to take query sessions into consideration in constructing the binned datasets.
We would also like to develop a cost-based optimizer to determine results to display, as opposed to our current rule-based approach.
We would also like to improve the standard webworkers package for \emph{Node.js} to provide linear speedups through parallelism.
While \project{} currently supports one dimensional aggregations, extending it towards two dimensions~(heatmaps) is theoretically straightforward.
Other avenues for future work include speculative query execution to further speedup querying and user guidance~\cite{dimitriadou2014explore}.
We would also like to explore possibilities for integrating progressive refinement with online aggregation.
\newpage
\bibliographystyle{IEEEtran}
\bibliography{document}

\end{document}